# Representation Matters in Longitudinal Affective Computing

Igor Matias, Maximilian Haas, Eric J. Daza,
Matthias Kliegel, and Katarzyna Wac

***Abstract*— Longitudinal, in-the-wild, wearable sensing yields day-level physiology, sleep, activity, and environmental streams, whereas affect and cognition are labeled only episodically (per waves). We recast this cadence mismatch as a temporal representation problem and compare three wave-level mappings from dense histories to sparse labels: levels (within-wave summaries), absolute drift (change across waves), and proportional drift. Using almost a year of data from 82 adults in the *Providemus alz* study, we model 21 affect and cognition outcomes. Day-scale signals are reduced to compact wave-level descriptors (central tendency, dispersion, and distributional shape) and learned with four regressors under two orthogonal evaluation axes—leave-one-subject-out and leave-one-wave-out. Performance is reported as scaled MAE using both mean and median across folds. Differences emerge: affective states are best predicted by wave-to-wave absolute drift, whereas cognitive performance aligns with within-wave levels, reflecting emotion dynamic theories. Across windowing features, shape descriptors (e.g., minima, kurtosis) carry more signal than simple means/medians. We contribute a representation triad for sparse-label modelling, a wave-level feature schema applicable on-device, and a dual-axis reporting practice that separates cross-participant generalization from temporal robustness. These results convert temporal representation from an implicit preprocessing step into an explicit, testable design choice for real-world affective-computing applications in brain health.**

*(Corresponding author: Igor Matias).*

Igor Matias is with the Quality of Technologies Lab and with the Cognitive Aging Lab of the University of Geneva, Switzerland (e-mail: igor.matias@unige.ch).

Maximilian Haas is with the Cognitive Aging Lab of the University of Geneva, Switzerland, and with the Faculty of Psychology of the UniDistance Suisse, Switzerland.

Eric J. Daza is with the Stats-of-1, Menlo Park, USA, and with the Boehringer Ingelheim Pharmaceuticals Inc., Ridgefield, USA.

Matthias Kliegel is with the Cognitive Aging Lab of the University of Geneva, Switzerland.

Katarzyna Wac is with the Quality of Life Technologies Lab of the University of Geneva, Switzerland

## I. INTRODUCTION

AFFECTIVE computing pursues computational methods that sense, model, and respond to human emotion in everyday life. Such emotion is traditionally assessed with gold-standard self-reports and *Ecological Momentary Assessments* (EMAs), while physiology, behavior, and other environmental signals are collected using sensors. A central challenge is to map those sparse and active affect labels to dense sensor streams when gathered in naturalistic settings (e.g., longitudinally, in everyday life, over months or years). In the lab, emotions can be annotated continuously; in the wild, gold-standard information arrives only episodically, while physiology and behavior are logged every day, hour, or even minute. In practice, cadence mismatches are often addressed either by increasing self-report burden—contrary to e-mental-health guidance [1], [2]—or by collapsing rich streams into coarse means or medians, discarding information about distributional shape and change. However, treating this frequency mismatch as an afterthought has encouraged ad-hoc windows and baseline heuristics that make models brittle and hard to compare [3], [4], [5]. Instead, in this article, we treat it as a representation choice—to be tested, reported, and replicated. This follows a long-standing theme in affective computing that representational assumptions shape inference, as in the case for treating affect ratings as ranks/ordinals rather than classes [6].

As one of the commonly used sensors in the study of human affect, consumer devices like smartwatches and fitness bands provide continuous, unobtrusive sensing of sleep, activity, and physiology at a population scale [7], [8], allowing for a deep exploration of everyday affective states. However, public-health frameworks increasingly group affective and cognitive functioning under an integrated brain-health umbrella that emphasizes ecological validity and equity [9], signaling the need to investigate both combined and separated. Literature in neuroscience shows that affect and cognition are deeply intertwined at the systems level [10]—sharing neural hubs, mutually modulating attention, memory, and control—so separating them analytically can be artificial in real-world behavior. In practice, the same wearable channels that index affective arousal (e.g., heart rate variability, electrodermal activity) also track cognitive load, and researchers have explicitly examined their interaction in naturalistic tasks (e.g., psychophysiological load effects on navigation [11]). Thus, in this research, we study not just affect but also cognition as a contrast class. Doing so helps exploring what is uniquely affective, while aligning with the broader brain-health framing that integrates affect and cognition [12].

Guided by emotion-dynamics theory—which characterizes affect as deviations around a personal set-point rather than static levels [13]—we ask a relevant question for affective computing: *Given wave-level affect assessments (i.e., labels longitudinally collected at multi-week or quarterly intervals) and day-level wearable streams, which temporal representation between both better generalizes across people and time and empowers modelling?* We study this in a naturalistic, multi-wave (multiple labels separated quarterly) cohort, and include cognition to isolate affect-specific results and to inform dual-tempo brain-health analytics (e.g., potential faster affective fluctuations in daily life versus slower cognitive variation).

Methodologically, we implement an end-to-end pipeline that (i) compresses day-long traces into wave-level descriptors suitable for privacy-preserving, on-device use; (ii) formalizes three wave-level temporal representations between sparse gold-standard labels and dense wearable data; and (iii) evaluates modelling under two generalization axes—*leave-one-subject-out* (LOSO–across people) and *leave-one-wave-out* (LOWO–across time)—so that within-participant interpolation is not mistaken for temporal model robustness.

Our contributions are methods-first. We present a representation-triad framework for sparse-label temporal representation that any *Intensive Longitudinal Data* (ILD) study can adopt; a compact wave-level feature schema that preserves informative aspects of sensor distributions while remaining edge-deployable [7], [8]; and LOSO/LOWO results that separate model generalization across participants from generalization across time. By treating this first as a representation problem, much like earlier work did for affect ratings, we propose a set of principled, reproducible design rules for fusing continuous/dense sensing with episodic labeling in real-world study of affect and cognition.

## II. RELATED WORK

Work in *IEEE Transactions on Affective Computing* (TAC) shows that representational assumptions materially alter inference—for example, encoding affect ratings as ordinal/ranked rather than nominal classes improves training fidelity and evaluation comparability [6]. Against this backdrop, empirical studies fall into two traditions with different assumptions regarding the most adequate temporal representation of gold-standard EMAs and questionnaires on affect (labels) and dense/continuous streams of data derived from devices like wearables.

Laboratory benchmarks of affective computing provide tightly segmented episodes with dense, contiguous labels and typically evaluate under LOSO (e.g., stress/amusement protocols such as WeSAD [14]). Such corpora are excellent for cross-participant generalization and intra-day affect recognition, but—by design—they avoid long inter-label gaps (e.g., weeks or months apart, studying long-term affect changes) and therefore do not test representations when labels recur, for example, weekly or quarterly.

In-the-wild/longitudinal mobile/wearable studies pair daily/weekly labels with continuous sensing, thereby revealing the cadence mismatch between sparse affect assessments and dense signals. Dense data temporal representation is most often operationalized by compressing signal streams via fixed windows [15] (e.g., “last day/week” mean or median, advantageous only when label recall period justifies it) or by increasing EMA frequency and burden, rather than by formalizing and comparing alternative representations [14], [16]. Recent longitudinal efforts advanced deployment-relevant evaluation by partitioning days into time slots for ambulatory ECG/accelerometry [17]; however, a direct temporal match between labels and signal streams continues not to be possible when using simple machine learning approaches like decision trees or linear models. Additionally, by only reporting LOSO model performance, researchers still do not address wave-level representations over longer horizons (labels separated by weeks or months), which are essential for long-term model validation and estimation of affective trajectories. Complementary surveys of real-life wearable emotion sensing report substantial heterogeneity in windowing, baselining, feature selection, and validation protocols, and explicitly call for standardized procedures to enable comparability [3]. Beyond TAC, cross-dataset evaluations in ubiquitous computing document poor reproducibility under divergent preprocessing/validation recipes [4], and broader mHealth reviews catalogue heterogeneous pipelines that impede synthesis [5].

Moving towards fulfilling those needs, we begin by basing our methodology on the core psychological and neurocognitive understanding of affect. A theory-driven strand—emotion dynamics—argues that affect is better characterized by instability and inertia around personal set-points than by absolute levels (e.g., changes in physical activity over months instead of the actual level of activity in a single month), implying that change-sensitive summaries may suit mood/affect phenomena. In contrast, and disentangling the cognition aspect of daily life, level-sensitive summaries (e.g., peak of physical activity minutes in a month) may better capture the theorized slower cognitive variation [18].

We start by adopting a standard within-wave windowing to reduce day-level streams into wave-level summaries, but we extend it by further exploring features beyond the mean and median. Then, our main contribution is orthogonal: we formalize a between-wave temporal representation that maps sensor histories to wave-level labels, comparing contemporaneous levels (within-wave) to drifts (between waves, absolute/proportional), asking which representation best supports generalization across people (LOSO) and across time (LOWO) for affective states and for cognition separately.

We therefore treat representation between labels and sensor streams as a testable problem at the wave level. This answers recent calls for principled, deployment-relevant evaluation [3], and avoids ad-hoc windowing by making temporal representation explicit, reportable, and replicable.

## III. Methods

### A. Dataset

*Population*

The Providemus alz project [19], of which this research is a part, is a two-year, longitudinal, observational study being conducted at the University of Geneva (Switzerland). Data collection began in March 2024 and will conclude in March 2026, divided into nine waves of active assessments (labels), each lasting approximately three months (labels repeated at the start of each wave, with wave 1 serving as the baseline). Volunteer subjects (participants) were eligible if they (i) were at least 45 years old at enrolment, (ii) resided primarily in Switzerland or France, (iii) were fluent in English, French, or Portuguese, (iv) used a smartphone daily, (v) could regularly wear and recharge a study-provided smartwatch, and (vi) were able to provide written informed consent. Individuals self-reporting any prior cognitive or brain health diagnosis were excluded. Ethical clearance was granted by the *Commission Cantonale d'Ethique de la Recherche sur l'être humain* with the number 2023-00975, and all participants provided written consent before enrolment.

Data from 82 participants were included in the analysis presented in this article. Their mean age at enrolment was 57.7 ± 8.5 years (range 45.5–77.6). Forty-nine (59.8%) were female. Most self-identified as White (76, 92.7%); the remainder identified as Asian (2, 2.4%), Hispanic/Latino (2), or another background (2). Participants reported a mean of 17.8 ± 4.7 years of education at baseline (range 6–37.7), and one-third (27, 32.9%) indicated a family history of dementia, while three were uncertain. Overall, this cohort provides a multilingual sample of mid- to late-adult individuals with mid- to high educational attainment.

*Instrumentation*

Participants installed the mQoL smartphone application[1] [19], [20] (Android / iOS) and paired their phone with a Withings Steel HR hybrid smartwatch (chosen for its battery life and accuracy [21], [22]). The phone app administered all active assessments (labels)—validated questionnaires (*patient-reported outcomes*, PROs) and gamified cognitive tasks (*performance-reported outcomes*, PerfROs)—and queried the Open Weather Map API for local atmospheric conditions (approximate GPS location only, due to privacy reasons, deleted after API reply and not stored for research). The smartwatch streamed passive signals (*technology-reported outcomes*, TechROs) on *heart rate* (HR), physical activity, and sleep; these were aggregated into daily summaries. Both devices operated continuously once set up, requiring only routine charging of the watch (around three weeks of battery life). All data were stored at the premises of the University of Geneva, in GDPR-compliant servers.

*Active Metrics*

Table I details the outcomes used to assess affective states and cognition. An earlier publication on the protocol [19] includes further details. A total of 21 labels (i.e., target variables) were measured at every wave. These were collected using the Buss-Durkee Hostility Inventory [23] (hostility), the Cognitive Telephone Screening Instrument [24] (inductive reasoning, long-term, prospective, short-term, and working memories, typing speed, and verbal fluency), the Hospital Anxiety and Depression Scale [25] (anxiety and depression), the Informant Questionnaire on Cognitive Decline in the Elderly [26] (cognitive decline), the Positive and Negative Affect Scale [27] (positive and negative affect), the Prospective and Retrospective Memory Questionnaire [28] (memory complaints), the Perceived Stress Scale [29] (stress), and the finger-tapping [30] (tapping speed), flanker [31] (inhibitory control), reaction time [32] (attention), and trail making [33] (cognitive flexibility, processing speed) tests.

Additional baseline controls included sex at birth, body mass index at age 40, *Cognitive Reserve Index Questionnaire* (CRIQ [34]), and *Mediterranean Diet Score* (MDS [35]) scores. At every wave, we also recorded self-reported alcohol intake, smoking status, years of education, cardiovascular or sleep-disorder diagnoses, head-injury history, and subjective age (with decade-based intervals as options, e.g., "31 to 40 years old"). These variables were used in subsequent modelling.

*Passive Metrics*

TechROs encompassed (i) physiological variables from the smartwatch, (ii) wear-time (percentage of the day for which any sensor data is collected, as defined in a previous publication [19], and follows the literature [36], [37]), and (iii) weather and air-quality information. A complete list of variables appears in Table I.

*Study Timeline*

*Providemus Alz* has nine assessment waves scheduled; this paper analyses the first four (Fig. 1). At the start of each wave, participants received app push notification-based and e-mail prompts and were asked to complete the battery of active tasks within 14 days, thus collecting the labels corresponding to that wave. However, tasks could have been split, reordered, or—even though discouraged—performed outside the 14 days. Passive sensing ran uninterrupted for the entire observation period, except for the moment when the devices were not used.

*Data Selection*

Four assessment periods (waves) were analyzed (97, 93, 90, and 33 days long, respectively). A participant's passive data in a specific wave was used only when at least 50 % of its days had a wear time of at least 10 hours, and when there was at least one atmospheric data point collected during the wave, as suggested by previous researchers [36], [37]. Zero-value HR samples were treated as missing. No data imputation was performed. Wave 1 had data from 57 different participants,

[1] http://mqol.unige.ch

TABLE I

LISTING OF ACTIVE AND PASSIVE DATA POINTS COLLECTED. DISTANCE RATE AND STEP FREQUENCY REPRESENT THE DISTANCE AND STEPS COVERED PER SECOND OF WEAR TIME IN A DAY, RESPECTIVELY.

| Active data labels per wave (21) | Passive data metrics (38, all TechROs), continuous across waves |
|---|---|
| **PROs**: anxiety, cognitive decline, depression, hostility, memory complaints, negative and positive affect, stress.<br><br>**PerfROs**: attention, cognitive flexibility, inductive reasoning, inhibitory control, long-term memory, processing speed, prospective memory, short-term memory, tapping speed, typing speed, verbal fluency (2 versions), working memory. | **Physical activity**: active calories burned, distance walking or running, distance rate, steps, step frequency.<br><br>**Sleep**: deep and light sleep duration, sleep efficiency and latency, sleep score, time spent in bed, total sleep time, wake-up count and latency, *wakefulness after sleep onset* (WASO).<br><br>**Sleeping HR**: mean, minimum, maximum.<br><br>**24h HR**: mean, standard deviation, median, minimum, maximum.<br><br>**Weather**: air temperature, atmospheric pressure, humidity, air temperature forecasted, temperature feeling<br><br>**Atmospheric pollutants**: ammonia ($NH_3$), carbon monoxide (CO), ozone ($O_3$), nitric oxide (NO), nitrogen dioxide ($NO_2$), *particulate matter* (PM) 2.5 and 10, sulfur dioxide ($SO_2$).<br><br>**Other**: wear time percentage. |

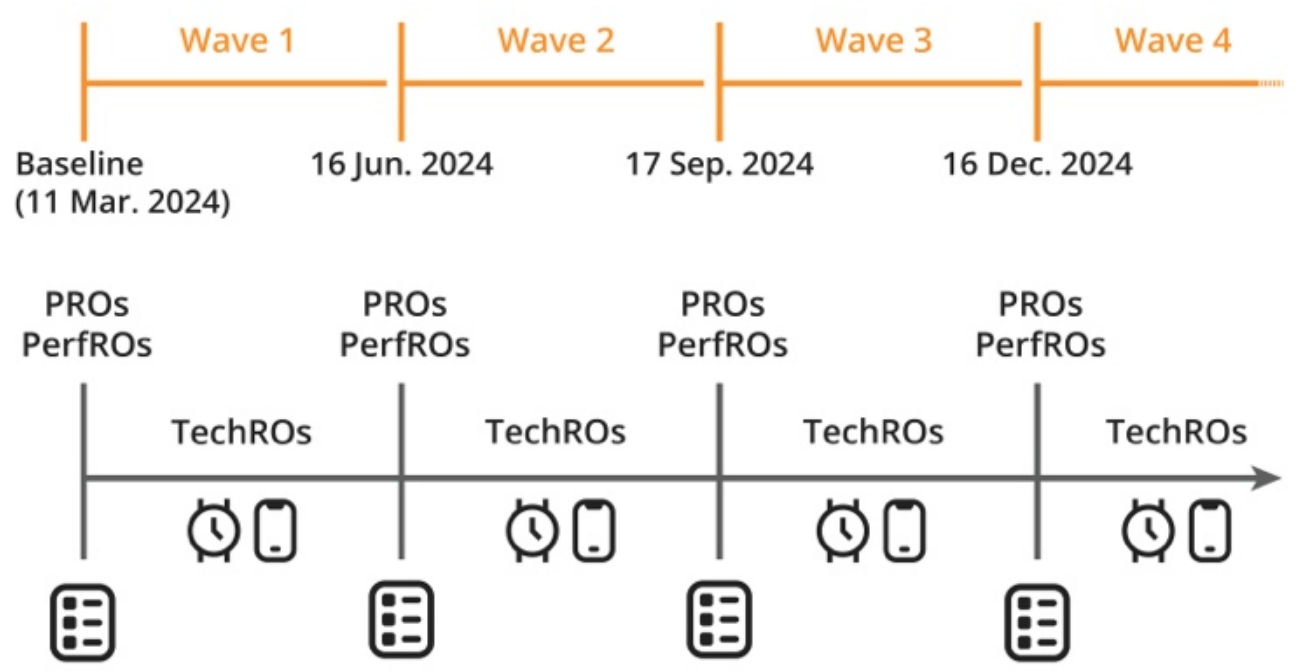

Fig. 1. Study timeline and data collection moments and tools.

wave 2 from 75, wave 3 from 77, and wave 4 from 73. Overall, the percentage of valid days per participant included in wave 1 was 85%, 96% for waves 2 and 3, and 94% for wave 4. The average percentage of the day with valid data across the included participants' data was 97% for waves 1, 2, and 3, and 92% for wave 4.

### *B. Temporal Representations and Wave-Level Summarization*

To link daily passive context (sensor streams) to periodic brain health outcomes (active data, labels), we first reduced both data streams to wave-level features, resulting in three distinct data representations: *absolute values* (Levels), *absolute difference of values* (ΔABS), and *percentage difference of values* (Δ%).

For the active data, the raw score obtained at each wave for every PRO and PerfRO served as the Levels. The other two representations required a pair of consecutive waves to be computed: ΔABS represented the score at wave $W + 1$ minus the score at wave $W$, and Δ% expressed the same change divided by the score at wave $W$. Because participants completed each wave at self-selected times, the exact number of days between consecutive assessments (Δdays) was recorded so that models could later adjust for unequal intervals in representations ΔABS and Δ%. We also computed two control variables per wave: the running mean of self-reported education years (since enrollment) and a subjective age gap obtained by subtracting the midpoint of the chosen age decade from the chronological age (negative values thus indicating participants who felt older than their age).

The passive data underwent a similar three-tier transformation. Over an entire wave's stream of data, Levels represented each of them with no changes, ΔABS subtracted from all of them the previous wave's median, and the Δ% was obtained by dividing the ΔABS values by the median in the previous wave.

Overall, the Levels representation served the purpose of modelling each participant's wave's data with no historical data for comparison, the ΔABS provided the change of one's wave's data compared with the wave before (a personalized three-month baselined approach), and the Δ% did similar but irrespective of one's range of values (a generalized three-month baseline, hence more generalizable population-wise). After having represented all data considering (or not) its history, we then summarized the wave-long passive data stream with eight descriptive statistics—*mean* (M), *standard deviation* (SD), *median* (Mdn), *inter-quartile range* (IQR), *minimum* (Min), *maximum* (Max), *skewness* (Skew), and *Kurtosis* (Kurt)—thereby condensing each sensor wave-long stream into a single value while retaining information on central tendency (M, Mdn), dispersion (SD, IQR), and distributional shape (Min, Max, Skew, Kurt). This was required according to the models later used in this research.

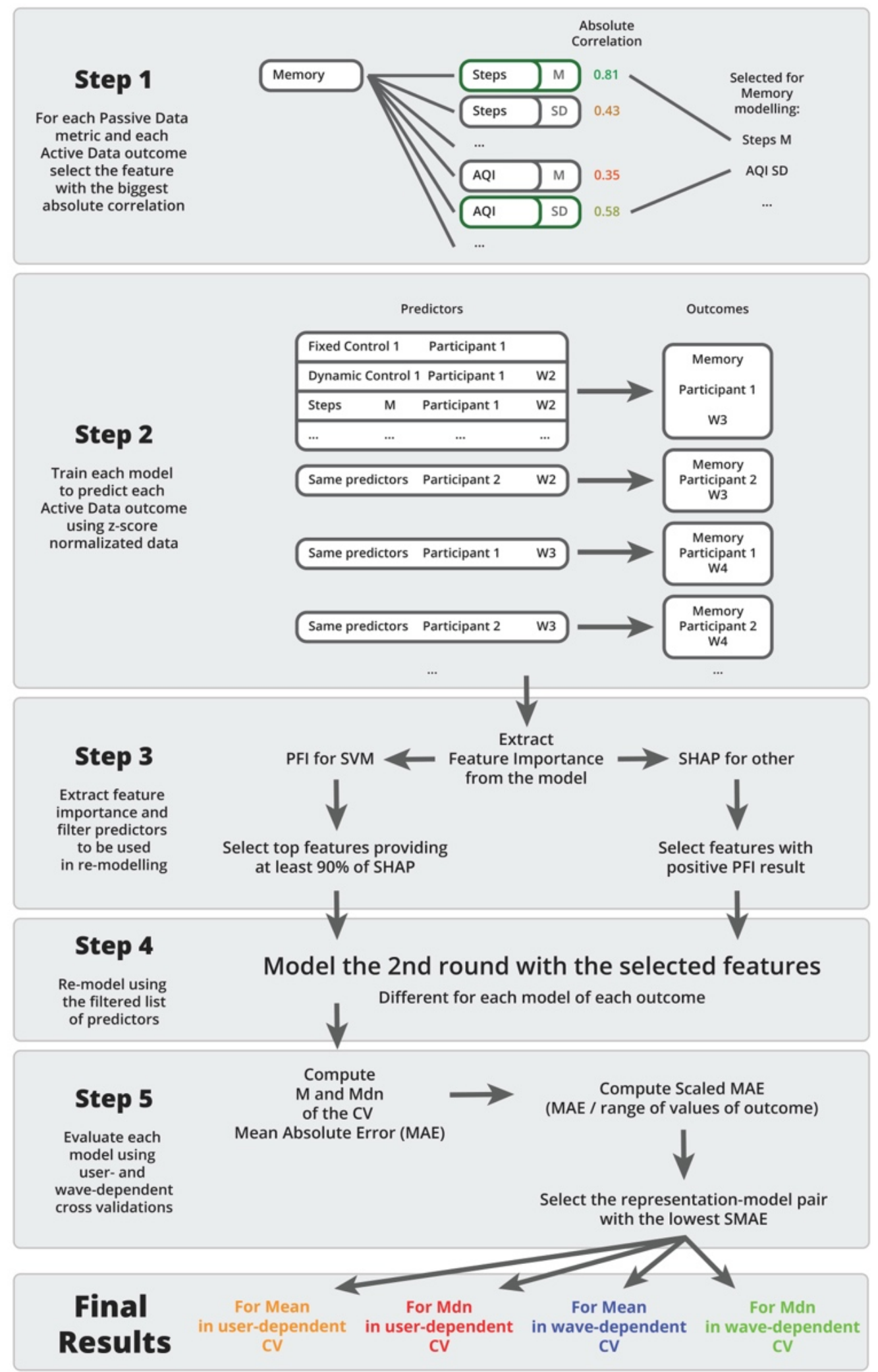

Fig. 2. Five-step diagram of the modelling pipeline.

Finally, given this research's goal of examining the power of passive data to predict actively collected brain health labels/outcomes, each passive data wave was merged with the active outcomes that followed, representing the relationship between behavioral and atmospheric exposures and the subsequent affective states and cognition. Thus, Levels of active points at baseline (wave 1) were discarded, as there was no passive data collected before them, and all the remaining passive data from wave $W$ were paired with active values at wave $W + 1$. Regarding ΔABS and Δ%, a similar pairing occurred; however, the first wave used was wave 2, as the median value of the preceding wave was required.

### *C. Modeling Pipeline*

Once the three wave–level representations were created, we applied a streamlined five–step pipeline (Fig. 2) to identify the most suitable representation option across the active outcomes (affect and cognition).

TABLE II
MODEL SETTINGS USED.

| Model | Settings |
|---|---|
| RF | Random state = 42 |
| LightGBM | Random state = 42<br>Number of estimators = 100<br>Learning rate = 0.1<br>Maximum depth = 6 |
| XGB | Random state = 42<br>Number of estimators = 100<br>Learning rate = 0.1<br>Maximum depth = 6 |
| SVM | RBF kernel, scikit-learn defaults |

First (step 1 in Fig. 2), for every outcome/label (21)–predictor feature (eight times 38) pair, we computed Pearson, Spearman, and Kendall partial correlations while controlling for chronological age, subjective-age difference, sex, and education. After z-score normalization on both ends of the correlation analysis, the strongest coefficient (the biggest absolute $r$) across the eight features in each coefficient defined a preliminary relevance score. Passive predictors for which the best $|r| < 10^{-3}$ were discarded. This step pruned redundant or uninformative features before training any learning algorithm.

Then (step 2), we trained four regressors selected for their complementary bias–variance profiles and transparent feature importances: *Random Forest* (RF), LightGBM, *XGBoost* (XGB), and *Support-Vector Regression* (SVM)—Table II includes the model settings used. The same set of controls used in the correlation analysis was used in the modelling phase; ΔABS and Δ% also included the Δdays control. Predictors and labels were z-score normalized to ensure fair comparison. Out of each model, feature importance was quantified (step 3). We used SHAP values for the three tree-based models and permutation importance for SVR. For each model, we retained the minimal subset of predictors that explained at least 90% of the cumulative SHAP or had a positive permutation score for SVR. This two-pass approach removed weak predictors that could inflate variance without sacrificing explanatory power. Each outcome was then re-modelled with its reduced predictor list (step 4).

Finally (step 5), to capture two real-world deployment scenarios, we used two CVs: LOSO to test generalization to unseen individuals, and LOWO to test temporal robustness within the same cohort. After each fold, we recorded the *scaled mean absolute error* (SMAE = MAE / original score/test range of output values, defined in the literature). Error distributions across CV runs could occasionally be skewed; so, we summarized performance with both the M (sensitive to large deviations) and the Mdn (robust to outliers).

For every outcome, we selected the representation–model pair that achieved the lowest SMAE under each CV scheme, as well as according to M and Mdn results. A comparison between regressors will be explored in another publication and is, therefore, out of the scope of this article.

## IV. RESULTS

### A. Feature Selection (Passive Data)

After selecting the features with the strongest correlation, we concluded that the most correlated—and therefore most used—metric was Kurt for models using Levels, IQR for models using ΔABS across waves, and Min for models using the Δ%. Conversely, the metrics less used across all representations were the M, SD, and Mdn.

That suggests that the effective way to summarize passive data from each wave, maximizing its potential predictive power, is to use metrics that account for its distribution and minima of values, rather than metrics that only inform the central tendency. More specifically, Levels tend to communicate more information through the presence (or absence) of extreme values (Kurt), ΔABS provides more information from the range of change over time (IQR), and Δ% highlights the importance of minimums in change rates over waves (Min). Putting together ΔABS and Δ%, such a difference in results may indicate that the standardization of change across participants makes the amplitude of variation less relevant, highlighting more the minimum fluctuations over time per participant.

### B. Data Merging and Cleaning

Once the most correlated features of each metric were selected to model each outcome, the passive and active data were combined. Given the missing values for some merged columns, the final number of input samples for each model varied. On average, Levels had 213.84 ± 3.18 samples, ΔABS had 130.60 ± 2.90, and Δ% had 48.63 ± 7.11. Although Δ% number of samples was substantially smaller compared with the other two representations, Levels and ΔABS still guarantee a valid comparison between absolute levels of sensor data and temporal drifts, which is the focus of our research.

### C. The Best Performing Data Representation

After performing the entire modeling pipeline, we could answer which data representation type achieves the lowest prediction error. Table III presents the complete results obtained, with the data representation leading to the lowest SMAE for each of the 21 active outcomes. One pattern was easily observed: PROs are always easier to predict when using ΔABS, whereas for PerfROs, this varies depending on the outcome.

Figures 3 and 4 show the frequency with which each data representation achieved the lowest SMAE at the end of the modeling pipeline, considering the M and the Mdn, respectively. Overall, across all 21 outcomes combined, ΔABS performed the most consistently, regardless of which CV we selected.

Interestingly, the values for LOWO remained unchanged between the M and Mdn SMAE cases, but they did change for the LOSO case (one case of Levels being the best performing in the M case changed to Δ% in the Mdn results). That may have been induced by the effect of participants to which the models could not generalize enough, thus showing an impact on the M SMAE while performing a user-dependent CV.

However, when analyzing the results for affective states and cognition outcomes separately, we see a different scenario. While the ΔABS of values seems to be the option with the most considerable predictive power for all outcomes together, Figures 5 and 6 show that this is only true for affective state outcomes. For cognitive outcomes, Levels yielded the most accurate results. In fact, for affective state modelling, ΔABS and Δ% are rarely the most effective options. At the same time, for cognition, we can see a more even distribution of the results across the three representations. Also noteworthy is that, in the modeling of cognitive outcomes, only when performing LOWO could we find a clear most performing representation. For LOSO, Levels and ΔABS are tied in both the mean and median of the results.

## V. IMPLICATIONS FOR AFFECTIVE- AND COGNITIVE-COMPUTING RESEARCH

A persistent obstacle in affective computing and brain-health analytics is the temporal mismatch between the low-frequency gold-standard labels produced by questionnaires or laboratory tests and the high-frequency passive streams generated by wearables and smartphones. When investigators ignore this discrepancy, they must either inflate participant burden by sampling active outcomes more often or discard most of the sensor signal through coarse aggregation (e.g., weekly means or medians), thereby weakening statistical power. The temporal representation framework presented in this study—comprising the representation triad and an empirically ranked catalogue of sensor-derived features—offers a principled bridge between these two regimes of observation in affective computing. By deciding in advance whether an outcome of interest is affective or cognitive in nature, a researcher can select the representation family (within-wave levels versus wave-to-wave deltas) and the minimal set of windowing descriptors needed to maximize predictive yield while minimizing data volume.

The framework lends itself to immediate deployment in two complementary scenarios. For affective monitoring, change-sensitive descriptors such as the wave-to-wave delta kurtosis of HR can be calculated locally on the device and compared with a personalized threshold dependent on past periods or waves (long-term). Exceeding that threshold would trigger a just-in-time breathing exercise or other emotion-regulation prompt, allowing interventions to be delivered before self-reported stress escalates. For cognitive estimation, the same device can maintain a rolling baseline composed of stable level-based metrics—monthly IQR of nightly sleep efficiency and resting HR (short-term), for example. Only when that baseline drifts persistently downward would the system recommend a remote cognitive check, thereby enabling scalable, low-friction screening for incipient decline.

Taken together, these implications support the development of a dual-tempo architecture for digital phenotyping: day-scale algorithms that respond to rapid affective fluctuations, and

TABLE III
DATA REPRESENTATION THAT RESULTED IN THE SMALLEST SMAE FOR EACH ACTIVE OUTCOME, BASED ON THE MEAN AND MEDIAN OF BOTH LOSO AND LOWO VALIDATIONS.

| | | Participant-dependent CV (LOSO) | | Wave-dependent CV (LOWO) | |
|---|---|---|---|---|---|
| | **Outcome** | **Mean** | **Median** | **Mean** | **Median** |
| Affective state | Anxiety | ΔABS | ΔABS | ΔABS | ΔABS |
| | Depression | ΔABS | ΔABS | ΔABS | ΔABS |
| | Hostility | ΔABS | ΔABS | ΔABS | ΔABS |
| | Negative affect | ΔABS | ΔABS | ΔABS | ΔABS |
| | Positive affect | Δ% | ΔABS | ΔABS | ΔABS |
| | Stress | ΔABS | ΔABS | ΔABS | ΔABS |
| Cognition | Attention | Levels | Levels | Levels | Levels |
| | Cognitive decline | ΔABS | ΔABS | ΔABS | ΔABS |
| | Cognitive flexibility | Levels | Levels | Levels | Levels |
| | Inductive reasoning | Levels | Levels | Levels | Levels |
| | Inhibitory control | Levels | Levels | Levels | Levels |
| | Long-term memory | Levels | Levels | Levels | Levels |
| | Memory | ΔABS | ΔABS | ΔABS | ΔABS |
| | Processing speed | Levels | Levels | Levels | Levels |
| | Prospective memory | ΔABS | ΔABS | Levels | Levels |
| | Short-term memory | Levels | ΔABS | Levels | Levels |
| | Tapping speed | ΔABS | Δ% | ΔABS | ΔABS |
| | Typing speed | ΔABS | ΔABS | ΔABS | ΔABS |
| | Verbal fluency 1 | ΔABS | Δ% | Levels | Levels |
| | Verbal fluency 2 | Δ% | Δ% | Δ% | Δ% |
| | Working memory | ΔABS | ΔABS | Levels | Levels |

PRO = patient-reported outcome; PerfRO = performance-reported outcome. Levels = within-wave summaries; ΔABS = wave-to-wave absolute change; Δ% = proportional change. LOSO = leave-one-subject-out; LOWO = leave-one-wave-out.
Cells list the representation with the lowest SMAE aggregated by mean or median of fold-wise errors under each CV scheme. Colors indicate outcome family (orange = PRO, green = PerfRO).

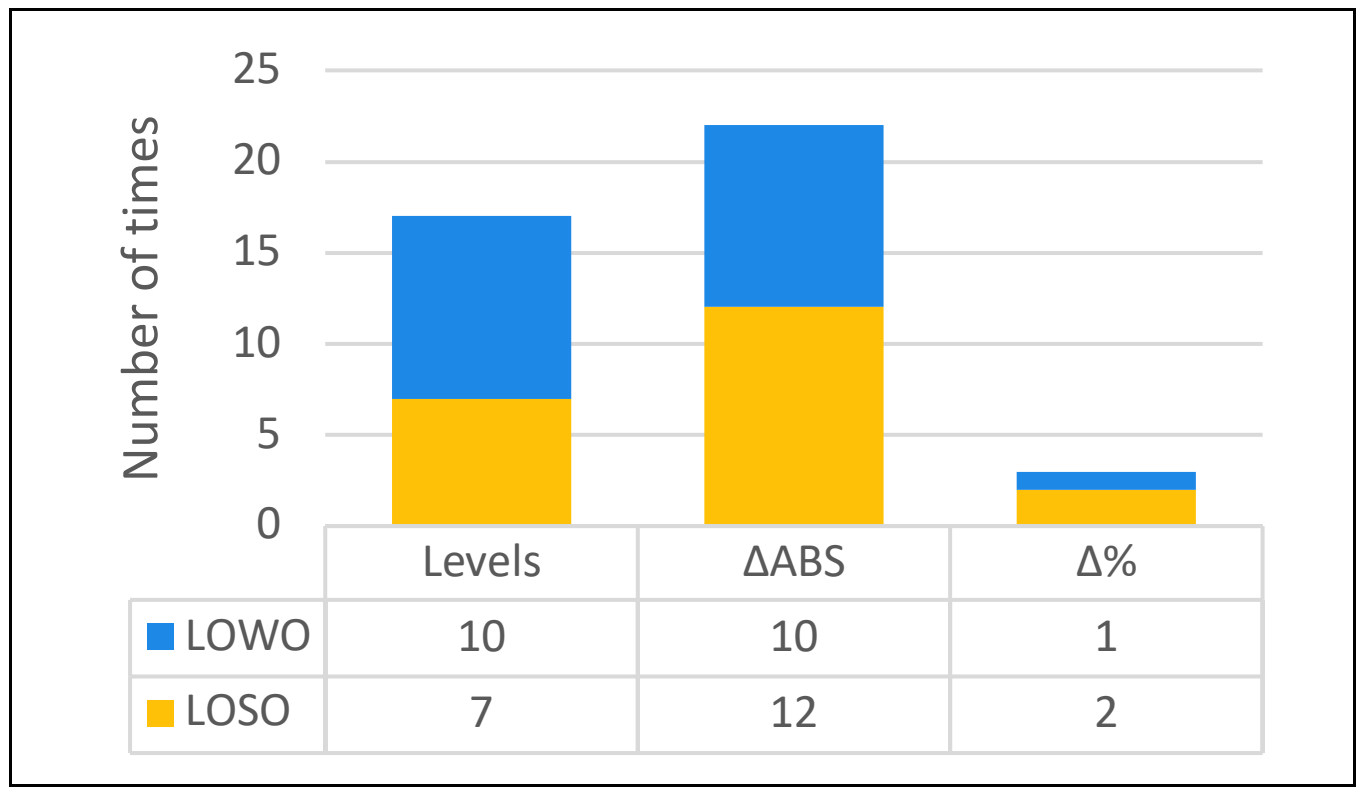

Fig. 3. Best representation type across all 21 active outcomes, considering the mean SMAE of LOSO and LOWO.

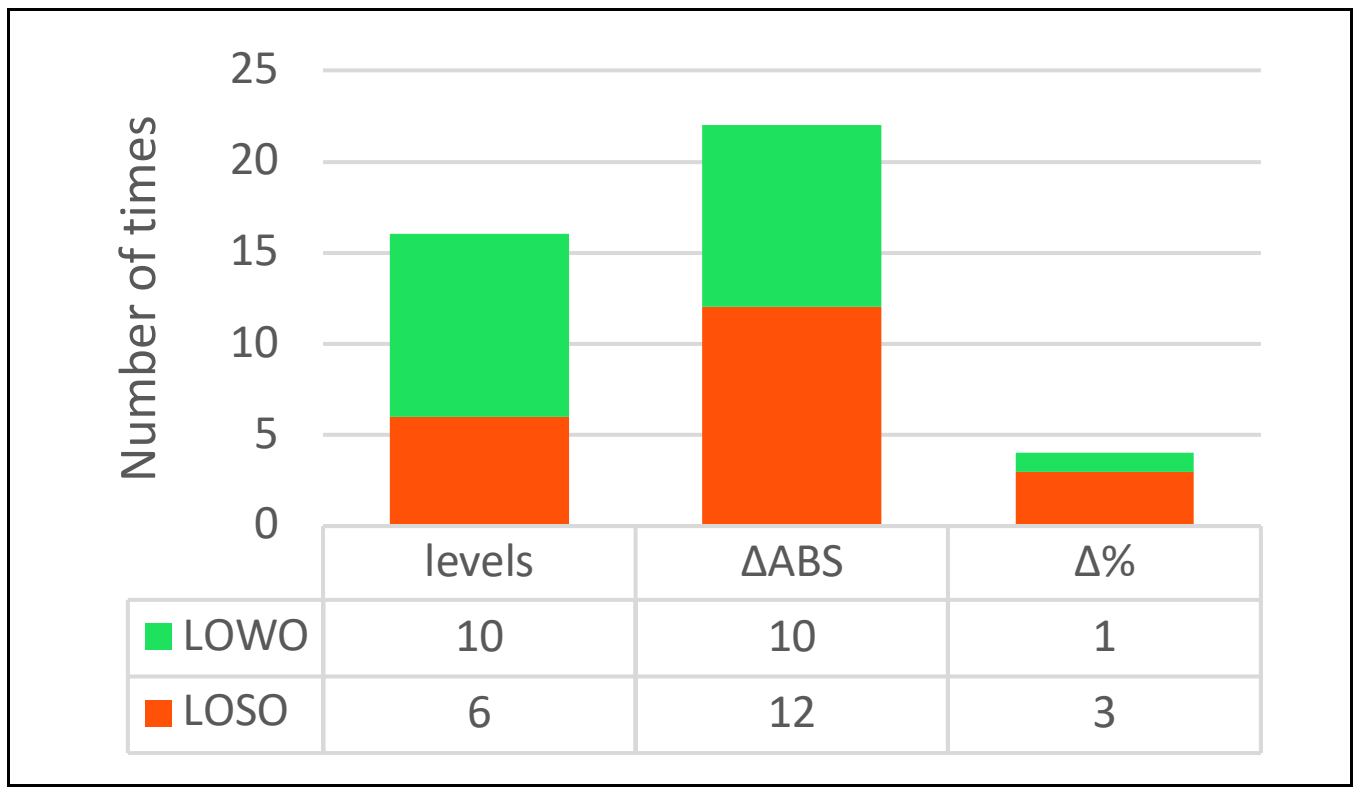

Fig. 4. Best representation type across all 21 active outcomes, considering the median SMAE of LOSO and LOWO.

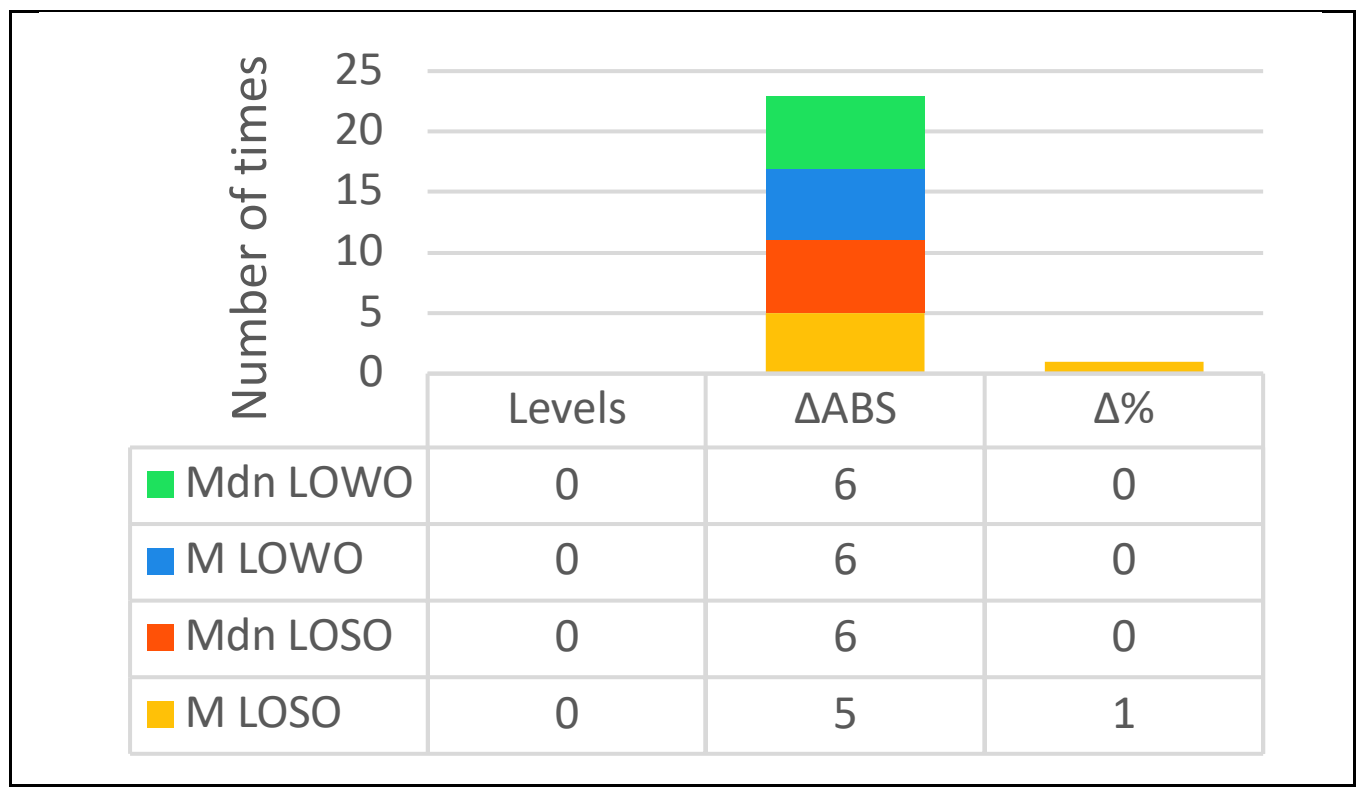

Fig. 5. Comparison of data representations' performance across affective state outcomes. The chart shows how often each data representation resulted in the lowest M and Mdn SMAE in LOSO and LOWO.

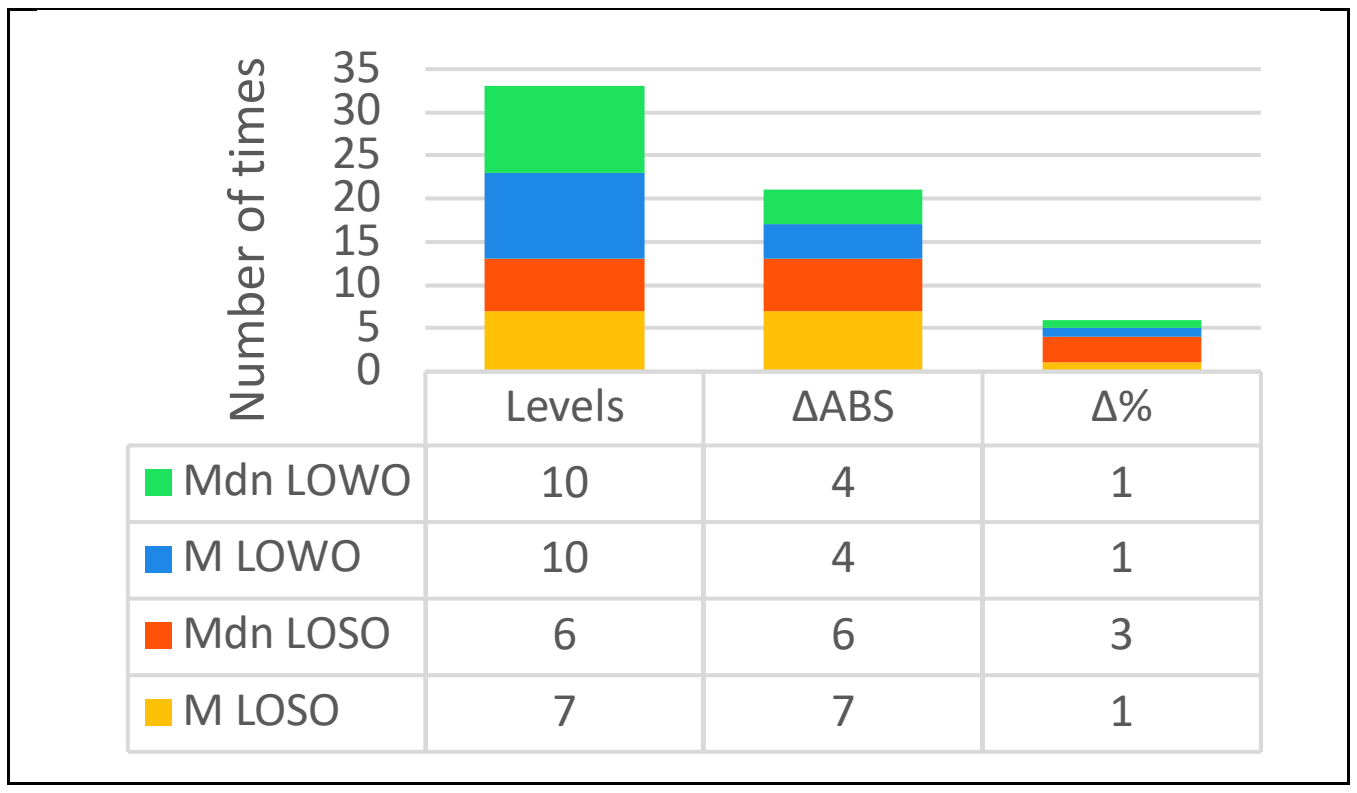

Fig. 6. Comparison of data representations' performance across cognition outcomes. The chart shows how often each data representation resulted in the lowest M and Mdn SMAE in LOSO and LOWO.

month-scale sentinels that monitor the gradual decline of cognitive capacity. Such an architecture accords with contemporary neuroscience, which views long-term cognitive integrity as a scaffold that modulates the stability of affective experience. By operationalizing this duality in a computationally lightweight and privacy-aware manner, the present work moves affective computing closer to the goal of proactive, continuous guardianship of brain health across the lifespan.

## VI. LIMITATIONS AND FUTURE PERSPECTIVES

Despite our systematic methods and sound results, some caveats might limit the generality of our conclusions. First, the cohort comprised cognitively healthy adults residing in Switzerland or France, predominantly Western, Educated, Industrialized, Rich, and Democratic [38]. Replication in socio-demographically diverse and clinically enriched samples is required before the findings can be considered population-invariant. Second, the present analysis covered only the first four quarterly waves; although this nearly 11-month span is atypically long for pervasive-sensing work, it likely remains insufficient to capture prodromal cognitive trajectories that unfold over years. Third, the wave-level summarization discarded finer temporal dynamics that might be crucial for detecting rapid affective fluctuations or short-lived cognitive changes. That may have obscured sub-day affect dynamics, but our choice reflects the temporal representation question rather than a claim of sufficiency for all affective tasks. Fourth, our cross-sectional feature-selection and correlation screens establish association, not causation, and are therefore prone to unmeasured confounding. Fifth, because data normalization was performed on the full dataset before cross-validation, some optimism in error estimates is possible. Finally, because control variables could be dropped between the first and second modeling phases, residual confounding cannot be ruled out.

Building on these constraints, two priority extensions follow. (i) Long-horizon tracking. The ongoing *Providemus alz* study will ultimately span two full years, more than doubling the current period, and enabling inquiry into brain health dynamics in healthy individuals and the effects of seasonality. (ii) Causal discovery. Incorporating invariant-risk minimization, temporal kernel methods, or other causal-inference frameworks will help separate mere covariance from influence (as previous research has focused on [39]), potentially guiding intervention design.

## VII. CONCLUSION

This work presents empirical evidence for addressing a deceptively foundational yet straightforward problem in affective computing: representing and aligning dense wearable streams with sparse self- and performance-based affective and cognitive reports in a manner that captures what truly drives human experience and changes over the lifespan. By evaluating three temporal representation strategies across 21 brain health labels in 82 adults over almost a year, we uncovered a potential double dissociation in two entangled brain-health components. Affective states were most accurately predicted by how a participant's physiology was changing—wave-to-wave deviations from their baseline carried the decisive signal. Cognitive performance scores, by contrast, depended on where that physiology sat: the absolute within-wave level of HR, sleep, etc.—an index of slower, trait-like biology (long-term personal baselines, not significantly changed over months). Two leave-one-out validations (subject- and wave-dependent) supported generalization both across individuals and over time.

Crucially, the delta-dominant result dovetails with dynamical theories of emotion, which posit that feelings arise when the organism detects deviations from homeostatic set-points (ΔABS) rather than absolute magnitudes of bodily arousal (Levels) [13]. It also echoes clinical findings that emotional inertia—the persistence or rapid escalation of physiological change—tracks mood-disorder severity better than raw levels do [40]. In computational terms, our results show that change-detection algorithms tuned to personal baselines can outperform traditional steady-state monitoring for predicting affective states. At the same time, despite being under the same brain-health umbrella, absolute thresholds remain more informative for cognition.

Beyond these theoretical insights, we contribute three main outputs: (i) Evidence that distributional “shape” features—kurtosis, IQR, and minima—outperform means and medians when day-long sensor traces are reduced to wave-level predictors. (ii) A representation triad that any intensive longitudinal study can adopt as a first-pass diagnostic. (iii) A dual cross-validation strategy and comparison that separates within-person interpolation from true out-of-sample generalization—essential for real-world deployment.

Together, these advances bridge the granularity representation gap between continuous passive sensing and episodic active affective and cognitive assessment, opening the door to near-real-time behavioral interventions powered by wearable data [41]. By showing precisely when change matters more than state—and when it does not—we move affective computing from passive observation toward proactive, personalized care, bringing the field a step closer to unobtrusive, always-on systems for monitoring affective states and cognitive health.

## Acknowledgment

A special acknowledgment is extended to all participants in the *Providemus alz* study, who have contributed their time and effort to make this research a reality. The first author also thanks all the co-authors and collaborators who helped build the tools and reasoning based on the collected data in this project, namely Dr. Alexandre De Masi, Dr. Paweł Prociow, and Dr. Clauirton De Siebra.

This research obtained funding from AGE-INT Swissuniversities, the *Centre Universitaire d'Informatique* of the University of Geneva, *Société Académique de Genève*, EU SHIELD (101156751), and by the Swiss National Centre of Competence in Research LIVES – Overcoming vulnerability: Life course perspectives, which is financed by the Swiss National Science Foundation (grant number: 51NF40-185901).

During the preparation of this work, the author(s) used ChatGPT to help draft Python scripts for data analysis and refine the writing of specific areas of the manuscript. After using this tool/service, the authors reviewed and edited the content as needed and take full responsibility for the content of the publication. All data processing and analysis were conducted locally using non-AI-powered tools. No participant's data was transmitted to any AI tool by any means.

## Declaration of interests

Eric J. Daza is a full-time employee of Boehringer Ingelheim and is the founder and chief editor of *Stats-of-1*. The other authors declare that they have no competing interests.

## Data sharing

Deidentified participant data underlying the findings of this study (e.g., wave-level features), along with the data dictionary, will be made available upon reasonable request to the first author. Access to the raw data is restricted due to ethical and privacy considerations, and a data access agreement will be required. Access to the raw data may be granted to researchers whose proposed use of the data has been approved by the study team.